\documentclass[aps,pra,preprint,superscriptaddress]{revtex4}

\usepackage{graphicx}
\usepackage{amssymb}
\small
\begin{document}

\title{Pull-in  control  due to  Casimir forces using external magnetic fields.}
\author{R. Esquivel-Sirvent}
\email[Corresponding author. Email:]{raul@fisica.unam.mx}
\affiliation{Instituto de F\'{\i}sica, Universidad Nacional Aut\'onoma
 de M\'exico, Apartado Postal 20-364, D.F. 01000,  M\'exico}
\author{M.  A. Palomino-Ovando}
\affiliation{Facultad de Ciencias Fisico-Matem\'aticas, Universidad Aut\'onoma de Puebla, Apartado Postal 5214, Puebla 72000, M\'exico}
\author{G. H. Cocoletzi}
\affiliation{Instituto de F\'{i}sica, Universidad Aut\'onoma de Puebla, Apartado  Postal J-48, Puebla 72570, M\'exico}

\date{\today}

\begin{abstract}

We present a theoretical calculation of the pull-in control in capacitive micro switches actuated by Casimir forces, using external magnetic fields. The external magnetic fields induces an optical anisotropy due to the excitation of magneto plasmons, that reduces the Casimir force. 
The calculations are performed in the Voigt configuration, and the results show that as the magnetic field increases the system becomes more stable.  The detachment length for a cantilever is also calculated for a cantilever, showing that it increases with increasing magnetic field.  At the pull-in separation,  the stiffness of the system decreases with increasing magnetic field.

\end{abstract}

\maketitle

With the decrease in dimensions of micro and nano electromechanical system (MEMS and NEMS)  into the nanometer range,  forces of quantum origin such as  the Casimir force  have become important.  The Casimir force can be understood as a retarded Van der Waals force between macroscopic bodies due to quantum vacuum fluctuations. This force has been measured precisely by several groups and for different configurations \cite{Lam97,Moh98,Cha01,Dec05,Ian04,george07}. The recent experimental and theoretical advances can be found in the  reports  by Bordag and  Klimchitskaya $et al.$ \cite{bordag}.  

The role of the Casimir force in MEMS  was first demonstrated theoretically by Serry and Maclay \cite{serry1,serry2} and experimentally by Buks \cite{roukes01,roukes012}, that showed  that this force in MEMS/NEMS devices can cause the movable parts to pull-in and adhere to each other.   The pull-in or snap down in MEMS and NEMS occurs when the magnitude of the Casimir force (or for that matter any attractive force) between the plates overcomes an elastic restitution force.  The pull-in in the presence of Casimir and Van der Waals forces has been 
 reviewed by several authors\cite{batra1, gusso,lin,delrio,esquivelapl,esquivelnjp,pinto}. Given that the pull-in is an undesirable effect,  its suppression in the electrostatic case and in the presence of dispersive forces has also been reported (see for example \cite{legrand,aerogel}).  
 
 In this paper we present a theoretical calculation of the effect of an external magnetic field  on the stability of a capacitive micro/nano switch, when the only force of attraction present is the Casimir force.  This external magnetic field can be used to control the pull-in dynamics since the Casimir force decreases with increasing magnetic field due to the excitation of magnetoplasmons \cite{cocol09}.

To test the idea consider the simple capacitive switch depicted in Figure 1. It consists of two parallel plates, one is fixed and the other plate is attached to a linear spring of elastic constant$\kappa$ and is allowed to move along the $z-axis$.  The plates are on the $x-y$ plane.  Given the frequency dependent dielectric function of the plates, the  Casimir force $F$ between two parallel plates separated a distance $L$ can be calculated using the Lifshitz formula.  For the purpose of this paper we use the reduction factor $\eta=F/F_0$ where $F_0=-\hbar c \pi/240 L^4$ is the force between perfect conductors. The reduction factor is given by  

 \begin{equation}
       \label{lifshitz}
       \eta=\frac{120 L^4 }{c \pi^{4}}\int_{0}^{\infty}Q dQ\int_{0}^{\infty}d\omega k (G^{s}+G^{p}),  
\end{equation}

where $G_s= (r_{1s}^{-1} r_{2s}^{-1} \exp{(-2  k L )}-1)^{-1}$ and $G_p=(r_{1s}^{-1} r_{2s}^{-1} \exp{(-2  k L
)}-1)^{-1}$. In these expressions, the factors
  $r_{p,s}$  are the reflection amplitudes for either $p$ or $s$ polarized light , $Q$ is the wavevector component along the
plates, $q=\omega/c$ and $k=\sqrt{q^2+Q^2}$. The above expressions are evaluated along the imaginary frequency axis $i \omega$, a usual procedure in Casimir force calculations \cite{bordag}. 
      
To be able to decrease the Casimir force via an external magnetic field, we have to change the reflectivities $r_p$ and $r_s$. This can be done using anisotropic media in the so-called Voigt configuration. In this configuration the magnetic field is parallel to the plates along the $x-axis$. 

Besides being experimentally more feasible,  in the Voigt configuration the Lifshitz formula Eq. (\ref{lifshitz}) can be used since there is no mode conversion of the reflected waves. In general for anisotropic media, an incident $p$ wave will reflect an $s$ and a $p$ wave and the same mode mixing happens for incident $s$ polarized waves.  In this case the Lifshitz formula has to be generalized using tensorial Green's functions and the reflectivities take a matrix form \cite{bruno}. 

Let us consider that the slabs are made of a semiconductor such as $InSb$, whose optical properties are well known \cite{manvir}.  In the presence of an external magnetic field, the components of the dielectric tensor are \cite{palik09} 

\begin{eqnarray}
\epsilon_{xx}&=&\epsilon_L\left[ 1-\frac{\omega_p^2}{\omega(\omega+i\gamma)} \right ], \nonumber \\
\epsilon_{yy}&=&\epsilon_L\left[ 1-\frac{(\omega+i\gamma)\omega_p^2}{\omega((\omega+i\gamma)^2-\omega_c^2)} \right ] ,\nonumber \\
\epsilon_{yz}&=&\epsilon_L\left[ \frac{i\omega_c\omega_p^2}{\omega((\omega+i\gamma)^2-\omega_c^2)} \right ], \nonumber \\
\end{eqnarray}
and  $\epsilon_{zz}=\epsilon_{yy}$ and $\epsilon_{zy}=-\epsilon_{yz}$.  The other components are equal to zero. 
In these equations $\epsilon_L$ is the background dielectric function, $\omega_p$ the plasma frequency, $\gamma$ the damping parameter and $\omega_c=eB_0/m^*c$ is the cyclotron frequency for carriers of charge $e$ and effective mass $m^*$. 
In the absence of the magnetic field, $\omega_c=0$ and the plates become isotropic. This anisotropy is induced by the excitation of magneto-plasmons.  

The reflectivities for $s$ and $p$ polarized waves in the Voigt configuration can be calculated using the surface impedance approach, as recently shown by \cite{cocol09}.   Using Eq.(\ref{lifshitz}) the Casimir force between the plates for different values of $\omega_c$ was calculated by , showing a reduction of  the  Casimir force as a function of magnetic field. \cite{cocol09}.  

The analysis of the stability of a capacitive switch, is well known (see for example\cite{pelesko02,batra1,zhao03} );  
  Given the total potential energy  of the system is $U=U_{elastic}+U_{Casimir}$, the critical point when the upper plate will jump to contact is 
 obtained from the conditions $\frac{dU}{dz}=0$ and  $\frac{d^2U}{dz^2}=0$. The first of these equations is simply the equilibrium of forces.  For example, for a force proportional to $\sim 1/r^n$, the pull-in distance is given by $z_{in}=(n/n+1)L_0$, where $L_0$ is the initial separation of the plates.  In the case of the Casimir force where $n=4$ the pull-in distance is $z_{in}=0.8 L_0$.   

 In our case,  using Eq.(1)  the equilibrium condition is  
 
 \begin{equation} 
\kappa (L_0-z)=\frac{\pi^2 \hbar c A}{240 z^4}\eta(z,\omega_c),
\label{force} 
\end{equation}
where $A$ is the area of the plates, and $\kappa$ is Hooke's constant.

Introducing the dimensionless quantities  $\bar{z}=z/L_0$ and $\Omega_c=\omega_c/\omega_p$,  the previous equation, Eq.(\ref{force}), is written as

\begin{equation}
\frac{(1-\bar{z})\bar{z}^4}{\eta(\bar{z},\Omega_c)}=\lambda,
\end{equation}
where  the parameter $\lambda$ is given by 
\begin{equation}
\lambda=\frac{F_0(L_0)}{\kappa L_0}. 
\end{equation}

 The plot of the parameter $\lambda$ as a function of plate separation $\bar{z}$ (bifurcation diagram) for different values of $\omega_c$ is shown in Figure (3).  As a reference we have plotted the bifurcation diagram for the Casimir force between  perfect conductors.  In this figure,  as the magnetic field increases the maximum value of $\lambda$ also increases.  For each value of the magnetic field, the upper part of the curve before the fold, corresponds to stable solutions. In all cases, the maximum value of $\lambda$ occurs at $\bar z=0.8$.  The inset shows the reduction factor for different values of the magnetic field at the pull-in separation. This calculation is based on the calculations of ref. \cite{cocol09}. 

A system that can be described by a simple-lumped system as that of Fig. (1) is a cantilever switch.  Consider a cantilever of rectangular cross section, of length $l$, width $w$,  thickness $t$  and Young modulus  $E$.  In this case the spring constant or stiffness is given by $\kappa=Ewt^3/4l^3$ \cite{handbook}.  From the definition of the bifurcation parameter,  we see that the pull-in stiffness can be calculated.  Similarly we can calculate the detachment length \cite{zhao03,guo04,zhao07} of the cantilever.  This is the maximum length of the cantilever that will not adhere to a substrate and is also determined from the values of $\lambda$ at pull-in. In Figure (3) we plot the stiffness and detachment length for different values of the cyclotron frequency.  As the magnetic field increases, the detachment length increases up to $60\%$  with respect to  the detachment length at zero magnetic field and the stiffness shows a $20\%$ decrease in its zero magnetic field value.

The reduction of the Casimir force due to the excitation of magnetoplasmons is used to control the pull-in or jump to contact in capacitive switches actuated by Casimir forces.  As the external magnetic field increases the force decreases allowing an increase of the bifurcation parameter  $\lambda$ at pull-in. From this value of $\lambda_{in}$ the pull-in stiffness and detachment length are calculated with increasing magnetic fields. In this paper we used dimensionless quantities, to present the results in general. Once a particular system and material is chosen, the magnetic fields needed to reach a value of $\Omega_c$ can be calculated. The additional parameter that allows us to change $\Omega_c$  in a given material is the plasma frequency,  that can vary in some orders of magnitude depending on the carrier concentration \cite{palik09}.  
\acknowledgements{Partial support provided by DGAPA-UNAM project No. IN-113208 and CONACyT  project No. 82474, VIEP-BUAP and SEP-BUAP-CA 191.}

\newpage

\begin{figure}
\includegraphics[width=8.5cm]{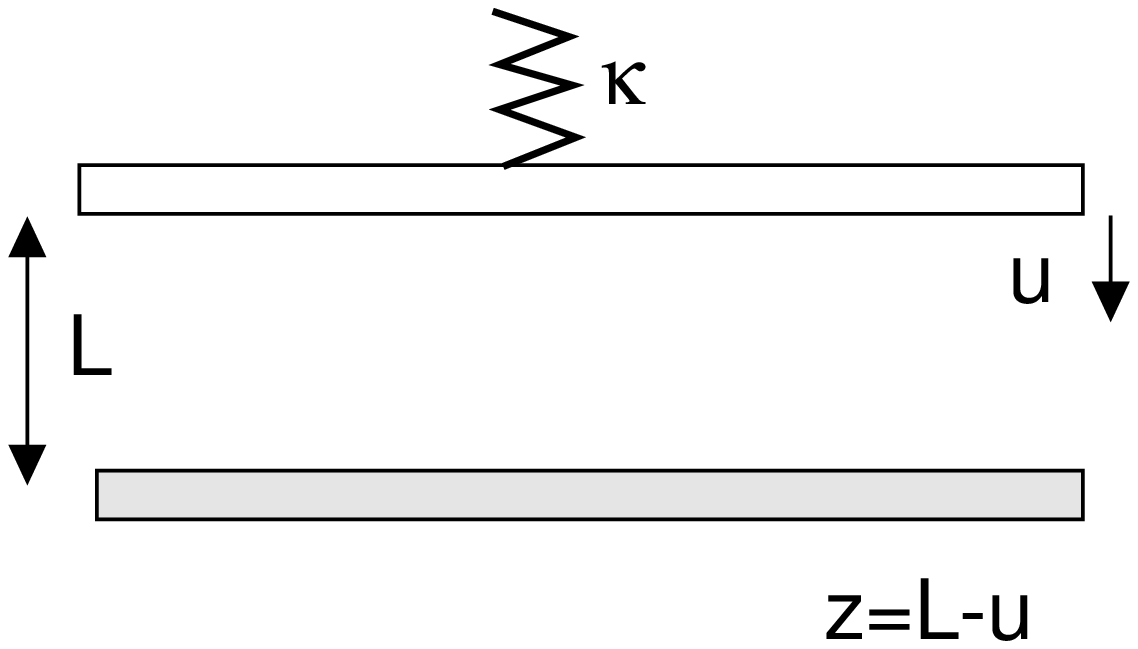} 
\caption{Simple one-degree of freedom system representation and coordinates used in our calculations.}
\label{fig9}
\end{figure}

\begin{figure}
\includegraphics[width=10cm]{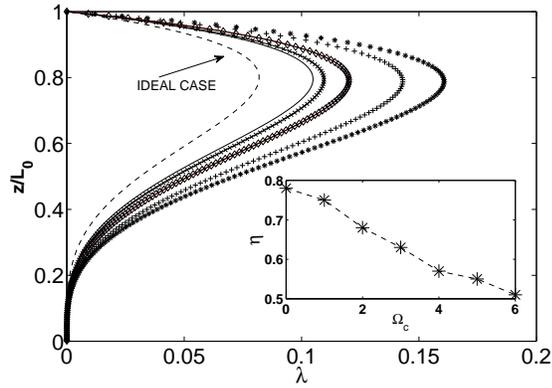}
\caption{Bifurcation diagram as a function of magnetic field. From right to left the curves correspond to 
the cyclotron frequencies $\Omega_c=6,5,2,1,0$. As a reference the bifurcation diagram for perfect conductors is also shown. The inset shows the reduction factor  as a function of $\Omega_c$. The reduction factor decreases with increasing magnetic field. This curve was calculated at a fixed separation between the plates $z=0.8 L_0$.  }
\label{fig9}
\end{figure}

\begin{figure}
\includegraphics[width=10cm]{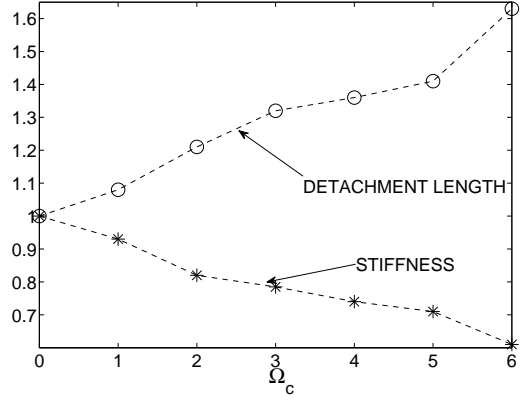}
\caption{The stiffness at the pull-in separation $\kappa/\kappa_0$ and the detachment length $l/l_0$ are plotted as function of the cyclotron frequency. The values of the stiffness and detachment length at zero magnetic field are $\kappa_0$ and $l_0$.  The detachment length can be increased up to a $60 \%$ as the magnetic field, and hence the cyclotron frequency, increases.}
\label{fig9}
\end{figure}

\end{document}